# A COMPARATIVE STUDY OF MORPHOLOGICAL CLASSIFICATIONS OF APM GALAXIES

A. Naim[1], O. Lahav[1], R. J. Buta[2], H. G. Corwin Jr.[3],
G. de Vaucouleurs[4], A. Dressler[5], J. P. Huchra[6], S. van den Bergh[7],
S. Raychaudhury[6], L. Sodré Jr.[8] & M. C. Storrie-Lombardi[1]

[1] Institute of Astronomy, Madingley Rd., Cambridge, CB3 0HA, U.K.
[2] Dept. of Physics and Astronomy, University of Alabama, Tuscaloosa, AL, 35487-0324 U.S.A
[3] California Institute of Technology, IPAC M/S 100, Pasadena, CA, 91125, U.S.A.
[4] Dept. of Astronomy, RLM 15.308, University of Texas, Austin, TX 78712-1083, U.S.A.
[5] Carnegie Observatories, 813 Santa Barbara street, Pasadena, CA, 91101-1292, U.S.A.
[6] Harvard-Smithsonian Center for Astrophysics, 60 Garden street, Cambridge, MA, 02138, U.S.A.
[7] Dominion Astrophysical Observatory, 5071 W. Saanich Rd., Victoria, BC, V8X 4M6, Canada
[8] Instituto Astronômico e Geofísico da Universidade de São Paulo, CP9638, 01065-970, São Paulo, Brazil

## ABSTRACT

We investigate the consistency of visual morphological classifications of galaxies by comparing classifications for 831 galaxies from six independent observers. The galaxies were classified on laser print copy images or on computer screen produced from scans with the Automated Plate Measuring (APM) machine. Classifications are compared using the Revised Hubble numerical type index T. We find that individual observers agree with one another with rms combined dispersions of between 1.3 and 2.3 type units, typically about 1.8 units. The dispersions tend to decrease slightly with increasing angular diameter and, in some cases, with increasing axial ratio ($b/a$). The agreement between independent observers is reasonably good but the scatter is non-negligible. In spite of the scatter the Revised Hubble T system can be used to train an automated galaxy classifier, e.g. an Artificial Neural Network, to handle the large number of galaxy images that are being compiled in the APM and other surveys.

## 1 INTRODUCTION

Since the introduction of the Hubble classification scheme (Hubble, 1926,1936) astronomers have been looking at ways to classify galaxies. Other systems were suggested, e.g. Mt. Wilson (Sandage 1961), Yerkes (Morgan 1958), Revised Hubble (de Vaucouleurs 1959), DDO (van den Bergh 1960a,b, 1976), and each has its special characteristics, but they all share Hubble's original notion that the sequence of morphologies attests to an underlying sequence of physical processes. This notion has been widely accepted for the past few decades, making morphological classification of large numbers of galaxies important for better modelling and understanding of galaxy structure and evolution. Examples include statistical relations which are specific to certain types of galaxies, e.g. the $D_n - \sigma$ relation for ellipticals (Lynden Bell *et al.* 1988), the Tully-Fisher relation for spirals (Tully & Fisher 1977) and the morphology-density relation (Hubble 1936, Dressler 1980).

Classification of galaxies is usually done by visual inspection of photographic plates. This is by no means an easy task, requiring skill and experience. It is also time consuming. The Third Reference Catalogue of Bright Galaxies (de Vaucouleurs *et al.*, 1991, hereafter RC3) contains nearly 18000 classifications. The ESO catalogue (Lauberts & Valentijn, 1989, hereafter ESO-LV) contains more than 15000 classifications. Both catalogues took several years to complete. However, in the APM (Automated Plate Measuring machine) survey (e.g. Maddox *et al.*, 1989) there are roughly $2 \times 10^6$ galaxies, and the expected yield of the Sloan Digital Sky Survey (Gunn *et al.*, in preparation) is over $10^7$ CCD images of galaxies. Clearly, such numbers of galaxies cannot be classified by humans. There is an obvious need for automated methods that will put the knowledge and experience of the human experts to use and produce very large samples of automatically classified galaxies. The first stage towards achieving this goal is creating a uniform, well-defined sample to be classified by human experts. This paper provides such a sample with six sets of independent human  classifications.

The outline of this paper is as follows : In § 2 we describe the galaxy sample. The Classification procedure is described in § 3. In § 4 we give the results of the statistical analyses carried out for the classifications and the discussion follows in § 5. The appendix gives a detailed listing of the sample and the classifications, including the derived mean types. The full original classifications appear in separate tables in the microfiche.

## 2   THE GALAXY SAMPLE

The galaxies were all taken from the APM Equatorial Catalogue of Galaxies (Raychaudhury *et al.*, in preparation), which is 98% complete for galaxies of magnitude $B \leq 17$ mag and $D \geq 0.5$ *arcmin*, covering most of sky between declination $-17°.5 < \delta < 2°.5$, and Galactic latitude $b \geq 20°$. The plates were IIIaJ (broad blue-green band) plates taken with the 48 in. UK Schmidt telescope at Siding Spring, Australia.

A diameter limited sample ($D > 1.2$ *arcmin* at an isophotal level of $24.5 mag/arcsec^2$) from 75 plates was compiled. The APM machine scans plates in strips 2.1 *arcmin* wide, and at the time this compilation was made the strips were analysed separately so large images were sometimes broken in two (depending on their orientation with respect to the scanning strip). For this reason the original list of APM-selected galaxies with $D > 1.2$ *arcmin* was augmented by galaxies from the PGC catalogue (Paturel *et al.* 1989) with $D > 1$ *arcmin*. After elimination of most duplicates and of galaxies that had severe contamination from overlapping stellar or galaxy images ($< 10\%$ of the sample) there remained 835 images (of which 831 were agreed by all observers to actually be images of galaxies). These were scanned from glass copies of the original plates, with a resolution of 1 *arcsec*, by the APM machine. The plates themselves, however, have a resolution of roughly 2 *arcsec* (due to observing conditions), which is therefore the limiting resolution of the digitised images. No plate matching was performed and no account was taken of possible brightness gradients within plates. A set of laser prints was prepared from the digitised images (hereafter the High Resolution Set, or HRS). A second set, of lower resolution prints, was also prepared by compressing the images to $128 \times 128$ pixels by averaging over groups of 4 pixels (or more, for the larger images) (hereafter the Low Resolution Set, or LRS). In figure 1 we show examples of four galaxies from the sample.

## 3   THE CLASSIFICATION PROCEDURE

Six of the authors classified the entire sample. The following initials will be used hereafter :

**RB** - R. Buta
**HC** - H. Corwin
**GV** - G. de Vaucouleurs
**AD** - A. Dressler
**JH** - J. Huchra
**vdB** - S. van den Bergh

Only vdB examined the images on a computer screen, while all other classifications were given for the hardcopies of the images. Six sets of classifications were given for the HRS. HC and GV also classified the LRS. RB, HC, GV, AD and JH gave their classifications on the Revised Hubble System as described in de Vaucouleurs (1959, 1963, RC3). Classifications by vdB followed the DDO System (van den Bergh, 1960a,b), and included Type as well as Luminosity Class (where applicable). HC also gave Luminosity Classes as part of his classification. Classification of each galaxy image took, e.g., GV, between 10 and 30 seconds.

RB, HC, GV and JH gave a full morphological description (e.g. SAB(rs)ab:) for each galaxy. AD gave just the T-type along the Revised Hubble sequence (table **??**). DDO classifications by vdB give Hubble type only to giant and supergiant galaxies, while dwarfs are assigned to broad classes such as E or S. For DDO classifications T-types were obtained by using the conversion tables in the Reference Catalogue of Bright Galaxies (de Vaucouleurs and de Vaucouleurs, 1964). In the cases where only a broad classification (e.g E, S) was given by vdB, no attempt was made to make a translation to T-type and such galaxies were considered as "not classified". Uncertainty symbols (i.e. ':' and '?') were used, but to significantly varying extents (see table 2).

For the statistical comparison, only T-types in the range -6 to 11 were considered as classifications. The very few cases in which galaxies were given T-types 90 (10) and 99 (Pec) were considered as "not classified", as were the few merger/interaction cases. For completeness, the RC3 classifications (where available) were added to give a total of 7 classifications. Appendix A contains a table of the numerical T-type classifications of all experts and the derived mean type (as explained below), and appendix B contains the full original classifications as given by each observer.

**Table 1.** T-types in the Revised Hubble System

| *cE* | *E0* | *E⁺* | *S0⁻* | *S0⁰* | *S0⁺* | *S0/a* | *Sa* | *Sab* | *Sb* | *Sbc* | *Sc* | *Scd* | *Sd* | *Sdm* | *Sm* | *Im* | *cI* |
|---|---|---|---|---|---|---|---|---|---|---|---|---|---|---|---|---|---|
| −6 | −5 | −4 | −3 | −2 | −1 | 0 | 1 | 2 | 3 | 4 | 5 | 6 | 7 | 8 | 9 | 10 | 11 |

**Table 2.** Basic Statistics of Classifications

|     | % classified | % uncertain | % with : | % with ? |
|-----|---|---|---|---|
| RB  | 91.8 | 33.6 | 19.5 | 14.0 |
| HC  | 97.3 | 76.5 | 47.9 | 28.5 |
| GV  | 56.8 | 32.6 | 11.5 | 21.0 |
| AD  | 97.7 | 2.0  | 1.9  | 0.1  |
| JH  | 98.9 | 1.6  | 0.0  | 1.6  |
| vdB | 65.7 | 10.5 | 8.9  | 1.6  |

## 4  STATISTICAL ANALYSIS

### 4.1  Mean Types

For each galaxy we calculated two mean types over the observers : One was a straight (unweighted) mean over all T-types given to that galaxy, while the other was a weighted mean, where (following Buta *et al.*, 1994) the weight given to a galaxy with no uncertainty symbol is 1, the weight for a ':' is 0.5 and the weight for a '?' is 0.25. We ended up with two lists of mean types for the whole sample.

In order to see whether there were any systematic differences between the two lists we employed a measure of variance between them :

$$(1) \qquad \sigma_{sw}^2 = \frac{1}{N_{gal}} \times \sum_{gal} (T_s - T_w)^2 \; ,$$

where $N_{gal}$ is the number of galaxies and $T_s, T_w$ are the straight-mean and weighted-mean types of each galaxy, respectively. For the two lists $\sigma_{sw} = 0.3$; The typical rms dispersions between any two observers were much higher, as will be discussed below, and therefore this value was taken to imply that the two lists are not significantly different from each other. Nevertheless, preference was given to the straight means since the varying extent of usage of the uncertainty symbols (see table 2 for details), meant that using weighted means could slightly bias the results.

### 4.2  Basic Statistics of the Classifications

The basic information regarding the classifications is shown in table 2. As can be seen, GV and vdB were the most conservative, classifying 57% and 66% of the images, respectively. The other three columns in the table state fractions out of the number of galaxies classified by each observer. The % uncertain column sums up the extent to which use was made of any uncertainty symbol, and the two columns to the right of it break this down to usage of ':' and of '?'.

vdB, AD and JH were more confident in their classifications, while RB, GV and in particular HC used the uncertainty symbols much more.

Figure 2, which depicts the classification histograms of each observer, shows a rough agreement between the classifications as well as some clear differences : There are defficiencies in Ellipticals (HC) and late type spirals (vdB); some distributions (GV, JH) peak at type 5, while others (RB, HC, vdB) peak at type 3 or (AD) type 4; There is a general tendency of vdB to classify mainly into "clean" types (e.g. E,S0,Sb) and avoid the intermediate types, as well as types 6–9.

In order to check the effect of image size and inclination on classifications we worked out the number of galaxies larger than a certain diameter (taken from the APM measurements) and of axial ratio (b/a) larger than a certain value (as calculated by our image reduction software). The results are shown in figure 3. It is clear that GV and vdB preferred the larger, less inclined images. This trend also exists, although weaker, in RB's classifications.

### 4.3  Comparison Between Classifications

#### 4.3.1  T-Type Statistics

Figure 4 shows scatter plots of 8 pairs of observers. The points have been artificially spread around their true values so as to avoid them being printed one on top of the other. To get an idea on the degree of agreement between the classification sets we

**Table 3.** Classification Agreements

| | | | | |
|---|---|---|---|---|
| **757** | galaxies (91%) have $\geq$ | 2 | identical classifications |
| **459** | galaxies (55%) have $\geq$ | 3 | identical classifications |
| **183** | galaxies (22%) have $\geq$ | 4 | identical classifications |
| **40** | galaxies ( 5%) have $\geq$ | 5 | identical classifications |
| **8** | galaxies ( 1%) have | 6 | identical classifications |

In all, 354 galaxies received a classification from all six observers.

**Table 4.** RMS Dispersions between Pairs of Observers; Individual RMS Dispersions and the derived RC3 internal error are shown in the bottom two rows.

| | RB | HC | GV | AD | JH | vdB |
|---|---|---|---|---|---|---|
| RC3 | 2.1 | 2.0 | 1.8 | 2.2 | 2.1 | 2.3 |
| RB | | 1.3 | 1.6 | 1.7 | 1.8 | 1.7 |
| HC | | | 1.5 | 1.8 | 1.9 | 1.9 |
| GV | | | | 1.7 | 1.8 | 1.9 |
| AD | | | | | 2.1 | 1.8 |
| JH | | | | | | 2.0 |
| $\sigma_i$ | 1.0 | 1.1 | 1.2 | 1.4 | 1.5 | 1.4 |
| $\sigma_{RC3}$ | 1.9 | 1.8 | 1.3 | 1.8 | 1.6 | 1.9 |

made a cumulative count of the numbers of galaxies which received at least a certain number of identical classifications. This comparison is shown in table **??**. We also tried to isolate extremely discrepant classifications. For each galaxy in the sample we looked at all the classifications it received and picked those cases where the range of types given to it was more than 4 types wide. We found 90 such cases.

To quantify the agreement, the rms measure between observers $i$ and $j$ was calculated as :

$$(2) \qquad \sigma_{ij}^2 = \frac{1}{N_{gal}} \times \sum_{gal} (T_i - T_j)^2 \ ,$$

where the $N_{gal}$ is the number of galaxies both observers classified. Results are shown in table **??**.

We adopt a working assumption according to which the "internal" dispersion of each observer is independent of the "internal" dispersion of any other observer. One can then write :

$$(3) \qquad \sigma_{ij}^2 = \sigma_i^2 + \sigma_j^2 \ ,$$

where $\sigma_i$ is the individual dispersion of observer $i$. Then, by standard $\chi^2$ minimisation, one can deduce the individual dispersions. These too are shown in table **??**.

The following points need to be noted :

1. The rms measure suggests that the disagreement for any pair of observers is in the range of 1.3 to 2.3 types. The root-mean-square of the $\sigma_{ij}$'s given in the table over the 15 pairs is 1.8 types. When calculated only over the subset of 354 galaxies which were given a classification by all observers, we obtained a dispersion of 1.6 types.

2. In general, agreements with the RC3 classifications are weaker than agreements between any two observers. This is expected since RC3 classifications were made on a variety of plate materials, while all 6 observers looked at the same plate material in our sample.

3. The closest agreements are between RB, HC and GV, who worked together in the past (although RB and HC learned the system independently).

4. The individual dispersions range from 1.0 to 1.5 types.

5. The errors in RC3 vary considerably (Buta *et al.*, 1994). This is reflected in the rather large internal errors derived for RC3 in this sample, which covers an area of the sky for which RC3 does not have good sources of classifications.

Another way of looking at the agreement between classifications is to monitor the fraction of all galaxies whose classifications differ by no more than a certain number of types. This calculation is presented in figure 5 for six selected pairs of observers. All other pairs give similar results.

The degree of perfect match (i.e., to within n=0 types) is not very high for any pair. Agreements in excess of 80% are obtained only to within 2 types, and in most cases 90% and more are achieved only to within 3 types.

### 4.3.2 Dependence of Agreements on Diameters and Axial Ratios

To further examine the effect of diameter and axial ratio, we calculated the effect of these parameters on the agreement between pairs of observers. The results are shown in figure 6. The expected improvement in agreement as the limiting diameter goes up is evident in almost all pairs. There are few cases where the trend is not monotonous or even reversed, but even in these cases the effect is small and can be attributed to few less clear-cut cases rather than be considered a systematic trend.

**Table 5.** RMS Dispersions in Pairs for Early Type ($\bar{T}_{all} \leq 0$) and Late Type ($\bar{T}_{all} > 0$) Galaxies

| | Early | | | | | | Late | | | | | |
|---|---|---|---|---|---|---|---|---|---|---|---|---|
| | RB | HC | GV | AD | JH | vdB | RB | HC | GV | AD | JH | vdB |
| RC3 | 2.1 | 2.1 | 1.8 | 2.4 | 2.2 | 2.3 | 2.2 | 2.0 | 1.8 | 2.2 | 2.2 | 2.5 |
| RB | | 1.4 | 2.0 | 1.7 | 1.7 | 2.0 | | 1.3 | 1.5 | 1.7 | 1.9 | 1.7 |
| HC | | | 2.4 | 2.2 | 1.8 | 2.4 | | | 1.3 | 1.7 | 1.9 | 1.7 |
| GV | | | | 1.9 | 1.9 | 1.9 | | | | 1.7 | 1.8 | 1.9 |
| AD | | | | | 1.9 | 1.7 | | | | | 2.1 | 1.9 |
| JH | | | | | | 2.0 | | | | | | 2.0 |

A rather ambiguous picture emerges from the axial-ratio dependence. For most pairs there is a strong decrease in agreements as we go to lower axial-ratios. This reflects the fact that less information is available to the observer as the objects become more and more edge-on. On the other hand, a few of the pairs seem to agree *better* as the axial ratios go down. This may reflect a greater disagreement between these observers for early-type galaxies (which are gradually excluded as the axial ratio drops) than for late-type galaxies, and this is checked for below.

### 4.3.3  Dependence of Agreements on Type

One of the problems of the Hubble system is separating visually between the different kinds of early-type galaxies (e.g. van den Bergh, 1989). Since no photometric information (such as light profiles) was made available in the classification of this sample, it is interesting to check for the effect of type on the agreements between the classification sets. To do that we repeated the pair-comparisons for early type galaxies ($\bar{T}_{all} \leq 0$) and for late type galaxies ($\bar{T}_{all} > 0$) separately. The results are shown in table **??**. Over the 58 early-type galaxies which received 6 classifications the rms dispersion was $\sigma_{early} = 1.9$, while over the 296 late-type galaxies that received 6 classifications the rms dispersion was $\sigma_{late} = 1.5$. There is an overall trend for better agreement on late type galaxies, although not between all pairs of observers. Classifications by any human observer reflect one's visual perception as well as the morphological parameters one uses to define the separate types of galaxies. While detecting the exact reason for the overall improvement in the agreement over late-type galaxies compared with early-type galaxies may not be straight forward, its mere existence is of importance.

### 4.3.4  Comparison of Classifications at Different Resolutions

As mentioned above, HC and GV classified both the HRS and the LRS. To see the effect of the degradation of picture quality on the classifications we calculated the rms dispersion between the HRS and LRS classifications for each separately. Denoting the rms dispersion between the HRS and the LRS as $\sigma_{HL}$, we find that HC had $\sigma_{HL} = 0.9$ and that GV had $\sigma_{HL} = 1.4$. One needs to note, however, that GV classified the LRS first, after not classifying such amounts of galaxies for quite some time. The comparison of his HRS and LRS classifications is influenced by this fact as well as by the changing image quality from one set to the other. We then compared the rms dispersion between HC and GV for each of the sets, and found that while for the HRS $\sigma = 1.5$, for the LRS the dispersion was higher : $\sigma = 1.8$. The scatter indeed increases with degradation of picture quality.

## 4.4  Comparison of Luminosity Classifications

Luminosity classification follows the RDDO system (van den Bergh, 1976) and the stages range from I to V, with 4 intermediate values, thus amounting to 9 steps altogether. Luminosity class I corresponds to strong, well developed arms, while the V corresponds to very weak arms. We investigated the consistency of luminosity classification between the two observers who assigned such classifications to the sample : HC gave luminosity classes to 381 galaxies (46% of the sample), and vdB to 320 (38%). The number of galaxies for which both gave luminosity classes is 227 (27%). For the purpose of numerical comparison we translated the classes from Roman digits to arabic digits, representing the intermediate classes as half integer numbers. Therefore type 1 corresponds to I, 1.5 to I-II, etc. (in RC3, only integer values are used, so 1 corresponds to I, 2 to I-II, and 9 to V).

The rms deviation between HC and vdB in Luminosity Class for the 227 galaxies is $\sigma_{ij} = 0.63$, or slightly more than a single step in the 9-step scale. However, when the distributions of luminosity classes are plotted (figure 7) there is a marked difference between the two : HC tends to find many more galaxies in the mid-range values ($2 - 3$), while vdB has a more uniform spread of luminosity classes over most of the range $1 - 5$. Another apparent difference is the smooth distribution curve of HC's classifications, suggesting use of the intermediate half-integer classes is as common as that of the integer classes.

**Table 6.** Frequencies of other Characteristics; Two sources are quoted from the literature : S1 is the revised classification of 1528 galaxies (de Vaucouleurs 1963); S2 is a study of the Second Reference Catalogue of Bright Galaxies (de Vaucouleurs *et al.* 1976) by de Vaucouleurs and Buta (1980).

|        | % Unclassified | % A  | % AB | % B  | % Peculiar | % Ext. Ring | % r  | % rs | % s  |
|--------|----------------|------|------|------|------------|-------------|------|------|------|
| **RB** | 0.2            | 10.1 | 12.9 | 21.0 | 8.0        | 7.1         | 9.4  | 6.8  | 25.7 |
| **HC** | 0.7            | 7.5  | 9.5  | 26.4 | 21.3       | 14.2        | 5.6  | 15.9 | 13.0 |
| **GV** | 11.8           | 17.7 | 5.9  | 24.8 | 7.1        | 3.1         | 12.8 | 6.5  | 25.0 |
| **JH** | 0.7            | N/A  | 5.5  | 8.5  | 9.1        | 1.7         | 5.3  | N/A  | N/A  |
| **vdB**| 2.6            | 74.5 | 7.2  | 7.8  | 6.3        | 0.1         | N/A  | N/A  | N/A  |
| **S1** | N/A            | 26.2 | 20.9 | 28.8 | N/A        | N/A         | 15.3 | 20.8 | 31.7 |
| **S2** | N/A            | 10.9 | 12.2 | 16.9 | N/A        | N/A         | 8.5  | 11.8 | 19.6 |

vdB prefers the integer-classes, and this results in a less smooth curve (effectively having a local minimum at most of the intermediate half-integer values). We see that although there is a rough correlation between the two sets of classifications, the definition of luminosity classes is not clear-cut.

### 4.5    Comparison of Other Characteristics

Morphological features which are not covered in the T-type system are very important to understanding galaxies. The existence of a bar bears on the internal dynamics of the galaxy (e.g., Ostriker and Peebles 1973); Peculiar galaxies may tell us about mergers; Rings have been studied and catalogued (Buta 1995). We think it is therefore worth while to study these morphological features in the classification sets. All but AD's set contain some degree of morphological description. RB, HC and GV gave the fullest description. In table ?? we quote the frequencies of family qualifiers (ranging from A for non-barred through the intermediate AB to B for barred), external rings, and variety qualifiers (ranging from r for internal ring, through the intermediate rs to s for S-shaped galaxies). The "peculiarities" column includes all cases where some degree of peculiarity was noted. Also reported is the fraction of galaxies that received no morphological description whatsoever in that particular set (this number may be different from the number of galaxies for which no T-type was given (*cf.* table ??). In some cases there was a partial morphological description, but no definite T-type was given). Numbers quoted are fractions of all classified galaxies in each set, which were flagged for the relevant morphological characteristic. Every reference to external rings, internal rings, S-shapes and bars was counted regardless of uncertainties. In many cases the family or variety qualifiers were skipped altogether by the observer, due to lack of information (e.g. edge-on images). For this reason the numbers for family and variety qualifiers do not add up to 100%. Note that vdB defined as "SA" any lenticular or spiral galaxy in which he could not see a bar, whereas RB, HC and GV described a galaxy as "SA" only if they were convinced there was no bar in it.

There are close agreements between RB, HC and GV in some of the features, although HC finds more cases of peculiarities and external rings. JH agrees with RB and GV on peculiarities but finds less of the other characteristics. It is difficult to draw conclusions from this table, since uncertain cases were taken together with certain ones, and mixed morphologies (e.g. mixed bar, AB) were given the same weight as pure morphologies.

## 5    DISCUSSION

### 5.1    On the Nature of Galaxy Classification

One of the frequent criticisms of the Hubble Classification scheme is that it is a non-quantitative scheme and therefore an inherently ambiguous one (e.g. Mihalas & Binney, 1981). Although the criteria for classification are accepted generally by all observers, it is possible for each of them to give slightly different weights to various criteria. On the one hand, the existence of good agreements (to within about 2 types) between pairs of observers over the APM sample presented here strongly supports the notion of a sequence of morphological types. This means that there is indeed a morphological sequence of some sort. On the other hand, the fact that the agreement is not very high for any exact Revised Hubble type (in the range -6 to 11) seems to suggest that there is some "fuzziness" in the actual classifications, or that classifying galaxies into a one-dimensional sequence (the T-type) is not enough, and a higher dimensionality of classification-space is required (e.g. full Revised Hubble System or the DDO system).

One possible source of this apparent "fuzziness" may be the relatively low quality of the images (e.g. the original images were in many cases overexposed, burning out important information about the bulge, and the quality of the printouts may have degraded the images further). As discussed above, there is indeed scatter between classifications by the same observer when introduced with sets of different quality (HRS and LRS in this case). This means that image quality could account for some of the scatter in the classifications, though one would suspect not all of it.

However, the images in this sample originated in photographic plates. In many cases they suffer from saturation or under-exposure, and from the non-linear response of the plate. Therefore, if any ambiguities in applying a classification scheme to this sample arise from these problems, it may well be that this is a problem one has to live with, at least until large surveys are made available on CCDs (whose dynamical range is larger and whose response is more linear).

**Table 7.** Division into 5 Broad Classes and Frequency Distribution of Classifications into these 5 Classes, for the 6 observers as well as for the following sources : RC3 ($D > 1.5\ arcmin$), the ESO Catalogue (Lauberts & Valentijn, 1989), the UGC Catalogue (Nilson 1973), and the 1528 galaxies reclassified by GV (de Vaucouleurs 1963, designated S1 in the table). Note : the coding of types in the UGC catalogue has no type between Sc and Irr, and therefore we included Sc/Irr and Irr in the broad class 5).

| Broad Class | T values |
|---|---|
| **E** | T $\leq$ -3.5 |
| **S0** | -3.5 $<$ T $\leq$ 0.0 |
| **Sa+Sb** | 0.0 $<$ T $\leq$ 4.0 |
| **Sc+Sd** | 4.0 $<$ T $\leq$ 8.5 |
| **Irr** | 8.5 $<$ T |

| Class : | $E$ | $S0$ | $S_{a+b}$ | $S_{c+d}$ | $Sm, I$ |
|---|---|---|---|---|---|
| RB | 3.7 | 21.5 | 46.4 | 23.7 | 4.7 |
| HC | 1.2 | 20.9 | 50.1 | 25.0 | 2.8 |
| GV | 6.9 | 9.7 | 34.1 | 44.2 | 5.1 |
| AD | 6.6 | 11.6 | 50.4 | 29.5 | 1.8 |
| JH | 6.2 | 17.4 | 40.3 | 32.6 | 3.5 |
| vdB | 13.8 | 10.4 | 60.3 | 13.1 | 2.4 |
| | | | | | |
| RC3 | 8.0 | 14.0 | 31.6 | 36.7 | 9.7 |
| ESO | 8.9 | 16.3 | 46.1 | 21.7 | 7.0 |
| UGC | 8.5 | 14.5 | 31.4 | 38.5 | 7.0 |
| S1 | 13.0 | 21.5 | 31.8 | 25.3 | 8.4 |

## 5.2 On Automated Classification

Storrie-Lombardi *et al.* (1992) experimented with a 1 *arcmin* diameter-limited sample of 5217 galaxies from the ESO Catalogue of Galaxies (ESO-LV). They used 13 parameters and trained an *Artificial Neural Network* to classify galaxies into 5 broad classes : E, S0, Sa+Sb, Sc+Sd, Irr. The results were : 64% success in classifying into the same broad class, and 96% success in classifying to within one such class. An extension of that pilot study to the full scale of 16 types in the ESO sequence (Lahav *et al.*, in preparation) gives an rms dispersion of $\sigma = 2.1$ between the ESO classification and the resulting network classification. This is only slightly higher than the typical dispersion between two human observers. A typical "internal" scatter when two identical nets with different initial random weights are used is about 0.6.

In order to get a rough idea on how good or bad the value of 64% is, we similarly lumped together the 17 classes in the Revised Hubble System into five broad classes, as shown in table **??**. This cruder binning gives an indication of the general frequency of major types of galaxies, and may be of use to researchers doing other surveys and to theoreticians who are modelling galaxy formation.

We repeated the comparisons between classification sets, this time using only the five broad classes. The results are shown in figure 8. These results are remarkably similar to those obtained using the Neural Network. Admittedly, the results quoted refer to two different datasets. The ESO-LV sample presented to the network contained physical information (colour, surface brightness) which was unavailable for the APM sample. However, classification by eye is usually done on photographic plates. The information on which the classification scheme is based is contained in the image of the galaxy on the plate and not in physical parameters, important though they may be. The ESO-LV parameters are, in this respect, *less adequate* for morphological classification than the actual APM pictures, and yet the Network managed to utilise them and come up with success rates that are comparable to the degree of agreement between human experts. This serves as a strong incentive for trying to classify the galaxies with a Neural Network based on nothing more than their digitised APM images. Indeed, work along these lines has been done (Naim *et al.*, 1995).

## 6 SUMMARY

We compared 6 independent classification sets for the same sample of 831 APM galaxies. We find good agreements between these sets but also detect non-negligible scatter. The overall rms dispersion of the observers is 1.8 types on the Revised Hubble T-system. Artificial Neural Networks give comparable results on a different sample of galaxies and we intend to apply them to the APM sample.

**ACKNOWLEDGEMENTS**

We thank the UKST unit of the Royal observatory of Edinburgh for the plate material, J. Lancashire and the APM group at RGO Cambridge for scanning support, M. Irwin, S. Maddox and D. Lynden-Bell for helpful discussions. JH was supported in part by NASA/HST grant GO2684.08-87A. SR was supported by NASA grant NAS8-39073. LSJ thanks the financial support provided by FAPESP and CNPq, and the hospitality of the Institute of Astronomy, Cambridge and the Royal Greenwich Observatory.

## APPENDIX A

Table A1 contains, for each of the 835 images, its 1950.0 coordinates (in degrees), the diameter based on which it was originally selected (in arc minutes), and the T-type classifications given. Also included are the straight mean (Ts) and the weighted mean (Tw) types, as described in § 4. Numerical types outside the range $[-6, 11]$ have the following interpretations : 66 - duplicate image (seen elsewhere in this sample); 77 - not a galaxy (plate faults, etc.); 88 - merger; 90 - type I0; 99 - Peculiar; 999 - no classification given.

## APPENDIX B

Tables B1–B5 (in the microfiche) contain the full original classifications of RB, HC, GV, JH and vdB, in this order. These include all their remarks as well.